\documentclass[preprintnumbers,amsmath,amssymb]{revtex4}

\usepackage{amsmath}
\usepackage{graphicx}
\usepackage{dcolumn}
\usepackage{bm}
\usepackage[usenames]{color}
\usepackage{multirow}

\usepackage{amsmath}
\usepackage{graphicx}
\usepackage{bm}

\newcommand{\be}{\begin{equation}}
\newcommand{\ee}{\end{equation}}
\newcommand{\ben}{\begin{eqnarray}}
\newcommand{\een}{\end{eqnarray}}
\newcommand{\ra}{\rangle}

\begin{document}

\preprint{draft version}

\title{The Relationship Between Entanglement, Energy,
and Level Degeneracy in Two-Electrons Systems}

\author{A. P. Majtey$^{1}$, A. R. Plastino$^{1,2}$
and J. S. Dehesa$^{1}$}

\address{
  $^{1}$Instituto Carlos I de F\'{\i}sica Te\'orica y
Computacional and Departamento de F\'{\i}sica At\'omica Molecular
y Nuclear, Universidad de Granada, Granada 18071, Spain.\\
  $^{2}$CREG-National University La Plata-Conicet, C.C. 727, 1900 La
Plata, Argentina.}

\begin{abstract}
The entanglement properties of two-electron atomic systems 
have been the subject of considerable research activity in recent years.
These studies are still somewhat fragmentary,  focusing on numerical computations
on particular states of systems such as Helium, or on analytical studies of
model-systems such as the Moshinsky atom. Some general trends are beginning
to emerge from these studies: the amount of entanglement tends to increase 
with energy and, in the case of excited states, entanglement does not 
necessarily tend to zero in the limit of vanishing interaction between 
the two constituting particles. A physical explanation of these properties, 
shared by the different two-electrons models investigated so far, is still 
lacking. As a first step towards this goal we perform here, via a perturbative 
approach, an analysis of entanglement in two-electrons models that sheds new 
light on the physical origin of the aforementioned features and on their 
universal character.

\vskip 
0.5cm

\noindent
Pacs: 03.65.-w, 03.67.-a, 03.67.Mn

\end{abstract}


\maketitle

\section{Introduction}

Entanglement constitutes one of the most fundamental phenomena 
in Quantum Mechanics \cite{BCS07,BV09,J07,BZ06,AFOV08,TMB10}. 
Entangled states of composite quantum systems exhibit non-classical 
correlations that give rise to a rich variety of physical
phenomena of both fundamental and technological significance. Quantum
entanglement can be considered in two different and complementary ways.
On the one hand, entanglement can be viewed as a resource. The controlled
manipulation of entangled states is at the basis of several quantum
information technologies. On the other hand, entanglement can be regarded as
a fundamental ingredient for the physical characterization of natural quantum
systems such as, for instance, atoms and molecules (for a comprehensive
and up to date review on this subject see \cite{TMB10}).  These two points
of view are closely related to each other, although the latter is somehow
less developed than the former. Concerning the second of the abovementioned
approaches, several researchers have investigated in recent years the phenomenon
of entanglement in atomic physics
\cite{TMB10,AM04,CMS07,OS07,OS08,CSD08,PN09,YPD10,MPDK10,HF11,KO10,K11}.
This line of enquiry is contained within the more general program of
applying tools and concepts from information theory to the analysis
of atomic and molecular systems 
\cite{DB09,LRAAE09,GD03,NA07,NA06,GN05,LS2011,SL08,SL09,Esqui11}.
Most of the studies on entanglement in two-electron systems
focused on the properties of the concomitant ground states.
However, the entanglement features exhibited by excited states of
two-electron atomic systems have also been explored \cite{YPD10,MPDK10}.
In this regard, the most detailed results have been obtained from analytical
investigations of the entanglement properties of exactly soluble models, specially
the Moshinsky one \cite{YPD10}. The behaviour of these soluble models is consistent with
some (partial) results yielded by numerical explorations of entanglement
in Helium based on high quality, state-of-the-art wave functions.
Some general trends begin to emerge from these investigations. First, and
not surprisingly, entanglement is found to increase with the strength of the
inter-particle interaction. Second, entanglement also tends to increase with energy.
Finally, the entanglement of excited states does not necessarily vanish in the limit
of zero interaction. The last two properties are, perhaps, less intuitively clear
than the first one. In fact, in a recent comprehensive review article
on entanglement in atomic and molecular systems by Tichy et al. \cite{TMB10} 
it is said that ``{\it $\ldots$  the limit of
vanishing interaction strengths does not necessarily yield a
non-entangled state $\ldots $ it remains open whether
this discontinuity effect has to be considered an artefact of the
entanglement measures that are used, or whether a physical
explanation will be provided in future.}''. 

It is remarkable that the various two-electron models where entanglement 
has been studied so far share the basic qualitative features mentioned
above. This suggests that these features may constitute generic properties
of these kind of two-fermion models. In this regard, we share the opinion
expressed by Tichy et al. \cite{TMB10}: ``{\it  Due to the non-integrability
of any non-hydrogen-like atom, theoretical studies of multielectron
systems have, so far, mainly focused on exactly solvable model atoms. 
While such models differ strongly from real multielectron atoms as concerns 
the interelectronic interaction and the definition of the confining potential,
they allow insight in some qualitative features}.''

The aim of the present work is to clarify the origin of the aforementioned
properties. To this end we are going to consider
a perturbative approach to this problem, regarding the term in the Hamiltonian
describing the interaction between the two electrons as a small perturbation.
We shall show that the eigenvalue degeneracy of the unperturbed Hamiltonian 
(describing independent particles) plays a crucial role in explaining the 
entanglement features of the ``perturbed'' system.

\section{Quantum Entanglement in Systems of Two Identical Fermions}

Correlations between two identical fermions that are only due
to the antisymmetric nature of the two-particle state do not
contribute to the state's entanglement
\cite{ESBL02,GM04,GMW02,GM05,NV07,BPCP08,ZP10,BBB07,Olivo08}. 
The entanglement of
the two-fermion state is given by the quantum correlations
existing on top of these minimum ones.  For example, 
a two-fermions state of Slater rank one (that is,
a state whose wave function can be expressed, in terms of an 
appropriate single-partice orthonormal basis,  as one 
single Slater determinant) must be regarded as non-entangled.
There are deep, fundamental physical reasons for this. On the one hand,
the correlations exhibited by such states are not useful as a 
resource to perform non-classical information transmission 
or information processing tasks \cite{ESBL02}. On the other hand,
the non-entangled character of these states is consistent with 
the possibility of associating complete sets of properties to 
both parts of the composite system (see \cite{GM04,GMW02,GM05} 
for a detailed analysis of various aspects of this approach).



Two useful quantitative measures for the amount of entanglement of 
a pure state $|\psi\rangle$ of a system of two identical fermions 
are expressed in terms of (see \cite{PMD09} and references therein) 
the linear entropy,
\begin{equation} \label{entanglementSL}
\varepsilon_L(|\psi\rangle)=1-2Tr[\rho_r^2],
\end{equation}
and the von Neumann entropy
\begin{equation}\label{entanglementvN}
\varepsilon_{vN}(|\psi\rangle)= -Tr[\rho_r\ln\rho_r]-\ln 2,
\end{equation}
of the single-particle reduced density matrix $\rho_r$.
Notice that according to the entanglement measures given by Eqs.
(\ref{entanglementSL}) and (\ref{entanglementvN}) a pure state that
can be represented by a single Slater determinant has no
entanglement (that is, it is separable).

The fermionic entanglement measures (\ref{entanglementSL})
and (\ref{entanglementvN})
are closely related to the Schmidt decomposition of pure states
 of systems constituted by two identical fermions \cite{NV07}. 
For any pure state $|\psi \ra$ of two identical fermions it is
posible to find an orthonormal basis $\{ |i\ra, \,\,\, i=0,1,\ldots \}$ 
of the single-particle Hilbert space such that the state $|\psi \ra$ 
can be written as
\be \label{fschmidt}
|\psi\ra \, = \, \sum_i \, \sqrt{\frac{ \lambda_{i}}{2} } \,\,
\Bigl(
|2i\ra |2i+1\ra - |2i+1\ra |2i\ra
\Bigr),
\ee
where the Schmidt coefficients $\lambda_i$ verify $0\le
\lambda_i \le 1$ and $\sum_i \lambda_{i} =1$ (in the case of systems 
with a single-particle Hilbert space of finite dimension $N$, we 
assume that $N$ is even and that the sums on the index $i$ run 
from $i=0$ to $i=N/2$). Then one has that the entanglement measures 
(\ref{entanglementSL}) and (\ref{entanglementvN}) can be expressed 
in terms of the Schmidt coefficients of the state $|\psi\ra$ 
respectively as \cite{NV07,PMD09},
\be \label{fentschmidt}
\varepsilon_L(| \psi \rangle ) \, = \,
1 - \sum_{i} \lambda_{i}^2.
\ee
and
\be \label{vfentschmidt}
\varepsilon_{vN}(| \psi \rangle ) \, = \,
 - \sum_{i} \lambda_{i} \ln \lambda_{i}.
\ee
In the particular case of systems of two fermions with
a single-particle Hilbert space of dimension four, the 
quantity $2\varepsilon_L$ reduces to the entanglement
measure (usually referred to as squared concurrence)
studied in \cite{ESBL02} (see also \cite{BPCP08}).
The entanglement measure given by equations 
(\ref{entanglementSL}) and (\ref{fentschmidt}) has been 
recently applied to the analysis of various physical systems
or processes, including electron-electron scattering processes 
\cite{BBB07}, the study of entanglement-related aspects of 
quantum brachistochrone evolutions \cite{BPCP08}, and the 
entanglement properties of two-electron atomic models 
\cite{YPD10}. As a final remark on entanglement in 
fermionic system we mention that in the present work we deal
with the fermionic case of the concept of {\it entanglement 
between particles}. This is not the only possible conception
of entanglement in systems of identical particles. In particular,
there is an approach to the study of entanglement in systems of 
indistinguishable particles which focuses on the entanglement 
between different modes (see, for instance, \cite{TMB10} and 
references therein).

\section{Perturbative Approach}

Let us consider a system of two identical fermions (``electrons'')
governed by a Hamiltonian of the form $ H=H_0+\lambda H' $,
where the unperturbed Hamiltonian $H_0$ corresponds
to two independent (non-interacting)
particles, $\lambda H'$ describes the interaction between
the electrons, and $\lambda$ is a small parameter. When this 
system is treated perturbatively, the perturbative corrections
to the eigenenergies correspond to some ``fine structure''
sitting on top of the main  pattern due to the spectrum 
of $H_0$. It is plain that within this scenario the leading,
zeroth-order contribution to the energy spectrum is independent
of the detailed structure of the perturbation  $H'$. As we shall presently
see, the situation is completely different when, instead the energy, we
calculate the entanglement of the system's eigenstates. When the unperturbed
energy eigenvalues are degenerate the leading (zeroth-order) contribution
to the eigenfunction's entanglement does depend, in general,
on the details of the perturbation.

Let us consider an $m$-fold degenerate energy level of $H_0$,
with an associated set of $m$ orthonormal eigenstates
$|\psi_j\rangle, \,\, j=1, \ldots m$. Since $H_0$ describes
two non-interacting particles, the $m$ eigenstates
$|\psi_j\rangle$ can always be chosen to be Slater determinants written
in terms of a family of orthonormal single-particle
states $|\phi^{(1,2)}_j \rangle$, so that
$|\psi_j\rangle=(1/ \sqrt{2})(|\phi^{(1)}_j\rangle |\phi^{(2)}_j \rangle -
|\phi^{(2)}_j\rangle |\phi^{(1)}_j\rangle )$. All the members of the
 subspace ${\cal H}_s$ spanned by the states $|\psi_j\rangle$
are eigenstates of $H_0$ corresponding to the same eigenenergy.
That is, energywise they are all equivalent. However, the different
members of this subspace have, in general, different amounts of
entanglement. Typically, the interaction $H'$ will lift the
degeneracy of the degenerate energy level.
If we solve the eigenvalue problem corresponding
to the (perturbed) Hamiltonian $H$ and take the limit
$\lambda \rightarrow 0$, the perturbation $H^{\prime}$ will
``choose'' one particular basis $\{ |\psi'_k\rangle_{\lambda\to 0} \}$
among the infinite possible basis of ${\cal H}_s$. The states
constituting this special basis will in general be entangled.
These states are of the form 
$
|\psi'_k\rangle_{\lambda\to 0}=\sum_{j=1}^m c_{kj}|\psi_j\rangle,
$
and are determined (according to standard perturbation theory \cite{Desai10}) 
by the eigenvectors of the $m\times m$ matrix $\tilde H$ with elements
given by
$
{\tilde H}_{ij}=\langle\psi_i|H'|\psi_j\rangle.
$
It is then clear that in the limit $\lambda \rightarrow 0$ the eigenstates of 
$H$ will in general be entangled.

Let $\tilde m$ be the number of different single-particle states
within the family $\{|\phi^{(1,2)}_j \rangle, \, 1, \ldots, m \}$.
It is a quite typical behavior that $\tilde m$ tends to increase
with the degree of degeneracy $m$ of the energy levels of $H_0$
which, in turn, tends to increase with energy (that is, it tends to
increase as one considers higher excited states). 
This explains (at least in part) why the range of entanglement-values
available to the eigenstates $\{ |\psi'_k\rangle_{\lambda\to 0} \}$
tends to increase with energy. Indeed, the maximum amount of
entanglement (as measured by (\ref{entanglementvN}))
that can be achieved by a linear combination of Slater
determinants constructed from the single-particle states
$\{|\phi^{(1,2)}_j \rangle \}$ is
\begin{equation} \label{up}
\varepsilon = \ln \Omega,
\end{equation}
where $\Omega $ is the integer part of ${\tilde m}/2$.
Expression (\ref{up}) provides an upper bound for the
entanglement of the states $\{ |\psi'_k\rangle_{\lambda\to 0} \}$.

In the present work we are going to focus on the 
entanglement properties exhibited by the eigenstates  
$\{ |\psi'_k\rangle_{\lambda\to 0} \}$ and on the 
entanglement upper bound (\ref{up}). In this regard, our
perturbative approach is unusual, since we are focusing on
``zeroth-order properties''. However, it must be stressed that
the amounts of entanglement of the states $\{ |\psi'_k\rangle_{\lambda\to 0} \}$
 are in general finite quantities that do not vanish when 
$\lambda \rightarrow 0$ and, consequently, constitute dominant 
aspects  of the entanglement-related features characterizing
 the system.

\section{Two Interacting spin-$\frac{1}{2}$ Fermions 
in an External Confining Potential}

We apply now the previous considerations to a system consisting of
two interacting spin-$\frac{1}{2}$ fermions in an external confining 
potential $U(x)$. The interaction between the particles is
described by the potential function $V(x_1-x_2)$, with $V$ an
 even function. The Hamiltonian of this system is then,
\begin{equation}\label{GeneralGonzalez}
H=-\frac{1}{2}\frac{\partial^2}{\partial
x_1^2}-\frac{1}{2}\frac{\partial^2}{\partial
x_2^2} + U(x_1)  + U(x_2)  + \lambda V(x_1-x_2),
\end{equation}
where $x_1$ and $x_2$ are the coordinates of the two
particles. We use atomic units ($m=1$, $\hbar=1$).
A relevant instance of this system corresponds to
the case of harmonic confinement, $U(x) = \frac{1}{2}\omega^2x^2$,
where $\omega$ is the natural frequency of the external
harmonic field. This case includes the Mishonsky atom 
\cite{AM04,YPD10,M68}, where the interaction between the
particles is also harmonic, $\lambda V(x_1-x_2)= \frac{1}{2}\lambda
\omega^2 (x_1-x_2)^2$, with $\lambda\omega^2 \geq 0$ being
 the square of the natural frequency of the interaction harmonic 
field. The Moshinsky atom is an exactly soluble system
whose entanglement properties have been studied in detail.
The examples considered here indicate that some important
entanglement-related features of the Moshinsky model are also 
encountered in more general systems.

We now apply the formalism of perturbation theory to a system described 
by the Hamiltonian (\ref{GeneralGonzalez}) with harmonic confinement 
and a generic interaction $V$ between the particles. 
The unperturbed Hamiltonian is then,
\begin{equation}
H_0=-\frac{1}{2}\frac{\partial^2}{\partial
x_1^2}-\frac{1}{2}\frac{\partial^2}{\partial
x_2^2} + \frac{1}{2}\omega^2x_1^2
+ \frac{1}{2}\omega^2x_2^2,
\end{equation}
and the perturbation,
\begin{equation}
\lambda H'=\lambda V(x_1-x_2).
\end{equation}
When $\lambda=0$, the model
consists of two independent harmonic oscillators with the same
natural frequency. Let $|n\rangle$ ($n=0,1,2,...$) be the
eigenstates of each of these oscillators. Then, the kets
$|n,\pm\rangle$ constitute a single-particle orthonormal basis
(the signs $\pm$ correspond, in standard notation, to the spin
state of the spin-$\frac{1}{2}$ particle). The eigenstates of
$H_0$ are characterized by two quantum numbers $n_1$ and $n_2$
corresponding to the alluded pair of independent oscillators. The
corresponding eigenenergies depend only on the value of the sum
$n_1+n_2$ and are $m$-fold degenerate with $m=2(n_1+n_2)+1$
($m=2(n_1+n_2)+2$) if $n_1+n_2$ is even (odd). Assuming that
$n_1+n_2$ is odd, with $n_1=n_2-1$, and taking spin into
consideration, we can choose the following set of $m$
antisymmetric eigenstates (all with the same energy),
\begin{eqnarray} \label{emeslaters}
|\psi_1\rangle=&\frac{1}{\sqrt{2}}&(|n_1,+\rangle|n_2,+\rangle-|n_2,+\rangle|n_1,+\rangle)\nonumber\\
|\psi_2\rangle=&\frac{1}{\sqrt{2}}&(|n_1,+\rangle|n_2,-\rangle-|n_2,-\rangle|n_1,+\rangle)\nonumber\\
|\psi_3\rangle=&\frac{1}{\sqrt{2}}&(|n_1,-\rangle|n_2,+\rangle-|n_2,+\rangle|n_1,-\rangle)\nonumber\\
|\psi_4\rangle=&\frac{1}{\sqrt{2}}&(|n_1,-\rangle|n_2,-\rangle-|n_2,-\rangle|n_1,-\rangle)\nonumber\\
&\vdots\\
|\psi_{m-3}\rangle=&\frac{1}{\sqrt{2}}&(|0,+\rangle|n_2+n_1,+\rangle-|n_2+n_1,+\rangle|0,+\rangle)\nonumber\\
|\psi_{m-2}\rangle=&\frac{1}{\sqrt{2}}&(|0,+\rangle|n_2+n_1,-\rangle-|n_2+n_1,-\rangle|0,+\rangle)\nonumber\\
|\psi_{m-1}\rangle=&\frac{1}{\sqrt{2}}&(|0,-\rangle|n_2+n_1,+\rangle-|n_2+n_1,+\rangle|0,-\rangle)\nonumber\\
|\psi_{m}\rangle=&\frac{1}{\sqrt{2}}&(|0,-\rangle|n_2+n_1,-\rangle-|n_2+n_1,-\rangle|0,-\rangle)\nonumber,
\end{eqnarray}
which are represented by single Slater determinants and
consequently have zero entanglement. A similar set of separable
eigenstates of $H_0$ can be chosen when $n_1+n_2$ is even.

We consider now a harmonically confined two-fermion system
with an interaction potential given by a repulsive Dirac delta function,
\begin{equation}\label{delta}
\lambda V =\frac{1}{2} \lambda \delta(x_1-x_2),
\end{equation}
For the first excited energy level of $H_0$ ($n_1+n_2=1$) which is
four-fold degenerate we then have,
\begin{equation}
{\tilde H}=\frac{1}{4}\sqrt{\frac{\omega}{2\pi}}\begin{pmatrix}
  0 & 0 & 0 & 0 \\
  0 & 1 & -1 & 0 \\
  0 & -1 & 1 & 0 \\
  0 & 0 & 0 & 0
\end{pmatrix}.
\end{equation}
and the corresponding eigenvectors can be written as
\begin{eqnarray}\label{eigenvec4}
|\psi_1'\rangle&=&\frac{1}{\sqrt{2}}(-|\psi_2\rangle+|\psi_3\rangle)\nonumber\\
|\psi_2'\rangle&=&|\psi_4\rangle\nonumber\\
|\psi_3'\rangle&=&|\psi_1\rangle\nonumber\\
|\psi_4'\rangle&=&\frac{1}{\sqrt{2}}(|\psi_2\rangle+|\psi_3\rangle).
\end{eqnarray}
In the limit of vanishing interaction, $\lambda
\rightarrow 0$, the eigenstates corresponding to the first two
excited energy levels of the full Hamiltonian tend to the states
(\ref{eigenvec4}), which have the following amounts of
entanglement,
\begin{equation}
\varepsilon_L(|\psi_2'\rangle)=\varepsilon_L(|\psi_3'\rangle)=0
\end{equation}
\begin{equation}
\varepsilon_{vN}(|\psi_2'\rangle)=\varepsilon_{vN}(|\psi_3'\rangle)=0
\end{equation}
\begin{equation}
\varepsilon_L(|\psi_1'\rangle)=\varepsilon_L(|\psi_4'\rangle)=\frac{1}{2}
\end{equation}
\begin{equation}
\varepsilon_{vN}(|\psi_1'\rangle)=\varepsilon_{vN}(|\psi_4'\rangle)=ln 2
\end{equation}
It can be verified after some algebra that the eigenvectors 
(and the associated amounts of entanglement) obtained in this case
coincide with those corresponding (in the $\lambda \to 0$ limit of 
the two first excited energy levels) to an harmonic interaction. That is,
they are the same as those associated with the Moshinsky model. We thus 
see that the harmonic and the Dirac delta interactions lead, for particles 
confined by an external harmonic well, to the same entanglement behaviour 
of the first excited states in the limit of weak interaction.

 We now consider the case of generic external (one dimensional)
 potential $U(x)$ and interaction $V(x_1-x_2)$ (with $V$ an
 even function) so that
 $H_0=\sum_{i=1}^2\left(\frac{\partial^2}{\partial x_i^2} + U(x_i) \right)$
 and $\lambda H^{\prime}=\lambda V(x_1-x_2)$, which leads to
\begin{equation}
 {\tilde H}=\begin{pmatrix}
  a & 0 & 0 & 0 \\
  0 & b & c & 0 \\
  0 & c & b & 0 \\
  0 & 0 & 0 & a
\end{pmatrix},
\end{equation}
with
  \begin{eqnarray}
  a&=&\frac{1}{2}(\langle 0 |\langle 1|H'|0\rangle| 1\rangle-
  \langle 0|\langle 1|H'|1\rangle|0\rangle\nonumber-
  \langle 1|\langle 0|H'|0\rangle |1\rangle+\langle 1|\langle 0|H'|1\rangle |0\rangle)\nonumber\\
  b&=& \frac{1}{2}(\langle 0|\langle 1|H'|0\rangle| 1\rangle+\langle 1|\langle 0|H'|1\rangle |0\rangle)\nonumber\\
  c&=& -\frac{1}{2}(\langle 0 |\langle 1|H'|1\rangle |0\rangle+\langle 1|\langle 0|H'|0\rangle |1\rangle)
  \end{eqnarray}
where $|0\rangle$ and $|1\rangle$ are the ground and
first excited eigenstates corresponding to the external confining
potential $U(x)$. The eigenvectors of $\tilde H$ are,
\begin{eqnarray}
|\psi_1'\rangle&=&\frac{1}{\sqrt{2}}(-|\psi_2\rangle+|\psi_3\rangle)\nonumber\\
|\psi_2'\rangle&=&|\psi_4\rangle\nonumber\\
|\psi_3'\rangle&=&|\psi_1\rangle\nonumber\\
|\psi_4'\rangle&=&\frac{1}{\sqrt{2}}(|\psi_2\rangle+|\psi_3\rangle).
\end{eqnarray}
The values of entanglement exhibited by these states are
$\varepsilon_L=\varepsilon_{vN}=0$ associated to $|\psi_2'\rangle$ and
$|\psi_3'\rangle$ and $\varepsilon_L=\frac{1}{2}$, $\varepsilon_{vN}=\ln
2$ associated to $|\psi_1'\rangle$ and $|\psi_4'\rangle$. This result 
generalizes the previous one, as we have solved the problem for generic 
interactions and external confining potentials. Here it is possible
to obtain general results for an arbitrary confining potential $U(x)$
for the case where one has (in the limit of vanishing interaction)
one particle in the ground state and one particle in the first 
excited state of $U(x)$, because the degeneracy of the concomitant 
energy level (of the two-partile system) can 
be determined directly without knowing the detailed energy spectrum
associated with $U(x)$. On the other hand, the properties exhibited 
by states of higher excitation in the limit of vanishing interaction 
do depend (via the degeneracy appearing in this limit case) on the
detailed eigenenergies of the confining potential $U(x)$. Consequently,
the analysis of the limit of vanishing interaction can be performed 
only in a case-by-case way. In the next section we are going to consider 
higher excited states in the case of a generic interaction potential 
$V(x_1-x_2)$ and a harmonic confining potetial.

\section{Entanglement Upper Bound for Excited States in the Limit of Weak interaction}

\begin{figure}[hb]
\begin{center}
\vspace{0.5cm}
\includegraphics[scale=0.75,angle=0]{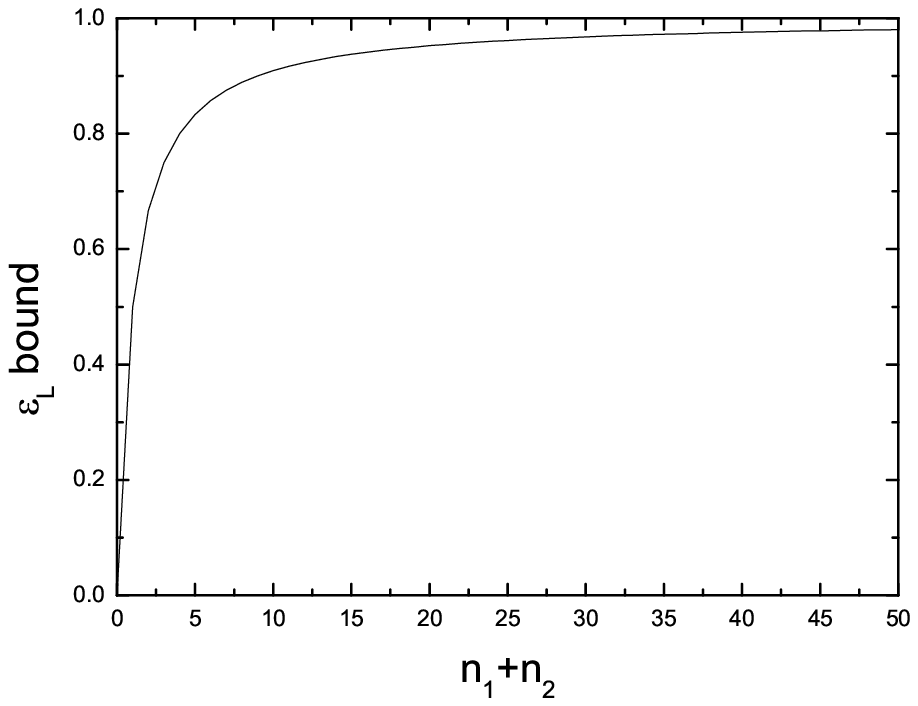}
\vspace{0.5cm}
\includegraphics[scale=0.75,angle=0]{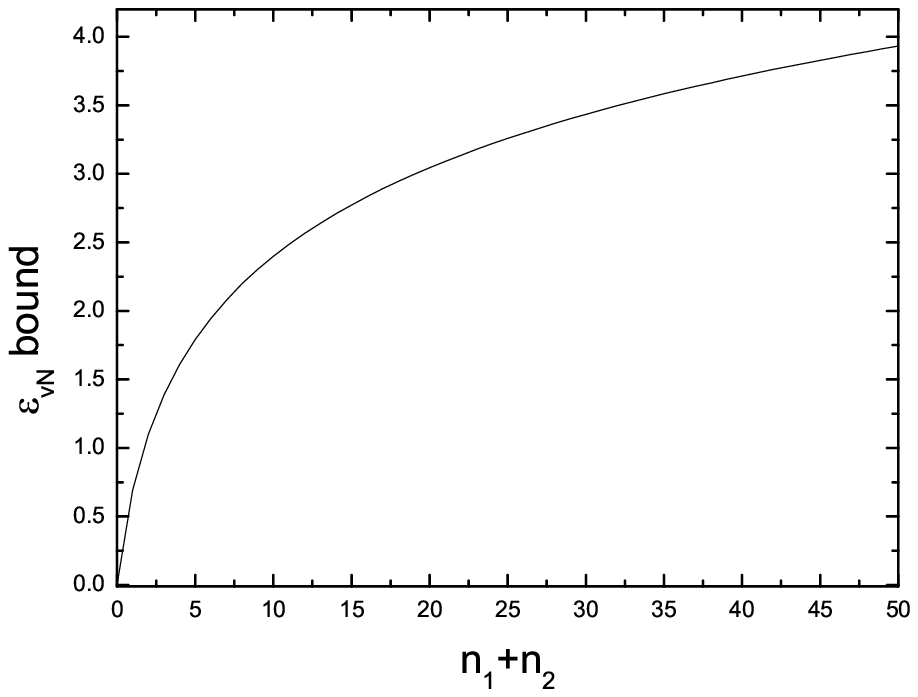}
\caption{Entanglement bound as a function of the sum of the
quantum numbers $n_1+n_2$. All depicted quantities are
dimensionless.\label{figu_1}}
\end{center}
\end{figure}

We consider now two spin-$\frac{1}{2}$ particles (in one
dimension) confined by an external harmonic potential and having a
generic interaction $\lambda V(x_1-x_2)$. We shall calculate
general upper bounds for the entanglement of the eigenstates of
this system in the limit $\lambda \rightarrow 0$. These bounds,
expressed in terms of the quantum numbers $n_1$ and $n_2$
characterizing the eigenfunctions of $H_0$, are,
\begin{equation} \label{up1}
\varepsilon_L(|n_1n_2\rangle)\leq\frac{n_1+n_2}{n_1+n_2+1}
\end{equation}
\begin{equation} \label{up2}
\varepsilon_{vN}(|n_1n_2\rangle)\leq ln(n_1+n_2+1)
\end{equation}
Equation (\ref{up2}) constitutes a particular instance,
corresponding to a harmonic confining potential,
of the general upper bound (\ref{up}).
In Fig. \ref{figu_1} we plot the  entanglement bounds
against $n_1+n_2$. These curves represent the maximum possible
entanglement compatible with those quantum numbers (the bounds do
not depend on the interaction and are, in this sense, universal).

In the particular case $n_1+n_2=2$, besides
 the above upper bound, we can calculate the 
exact amount of entanglement (in the limit 
$\lambda \rightarrow 0$) for an arbitrary 
interaction potential $V(x_1-x_2)$. This
case corresponds to a five-fold degenerate energy 
level of $H_0$ shared by a set of eigenvectors 
of the form (\ref{emeslaters}), with $m=5$. The 
generic matrix ${\tilde H}$ is 
of the form
\begin{equation}{\tilde H}=\begin{pmatrix}
  a & 0 & 0 & 0 & 0\\
  0 & b & c & 0 & d\\
  0 & c & b & 0 & -d\\
  0 & 0 & 0 & a & 0\\
  0 &d & -d & 0 & e
\end{pmatrix},
\end{equation}
with
\begin{equation}
 \begin{split}
  a=&\frac{1}{2}(\langle 0 |\langle 2|H'|0\rangle| 2\rangle-
 \langle 0|\langle 2|H'|2\rangle|0\rangle-\langle 2|\langle 0|H'|0\rangle |2\rangle
 +\langle 2|\langle 0|H'|2\rangle |0\rangle)\\
 b=&\frac{1}{2}(\langle 0 |\langle 2|H'|0\rangle| 2\rangle+\langle 2|\langle 0|H'|2\rangle|0\rangle)\\
 c=&-\frac{1}{2}(\langle 0 |\langle 2|H'|2\rangle| 0\rangle+\langle 2|\langle 0|H'|0\rangle|2\rangle)\\
 d=&\frac{1}{2}(\langle 0 |\langle 2|H'|1\rangle| 1\rangle+\langle 2|\langle 0|H'|1\rangle|1\rangle)\\
 e=&\langle 1 |\langle 1|H'|1\rangle|1\rangle.
 \end{split}
 \end{equation}

\begin{figure}[h]
\begin{center}
\vspace{0.5cm}
\includegraphics[scale=0.75,angle=0]{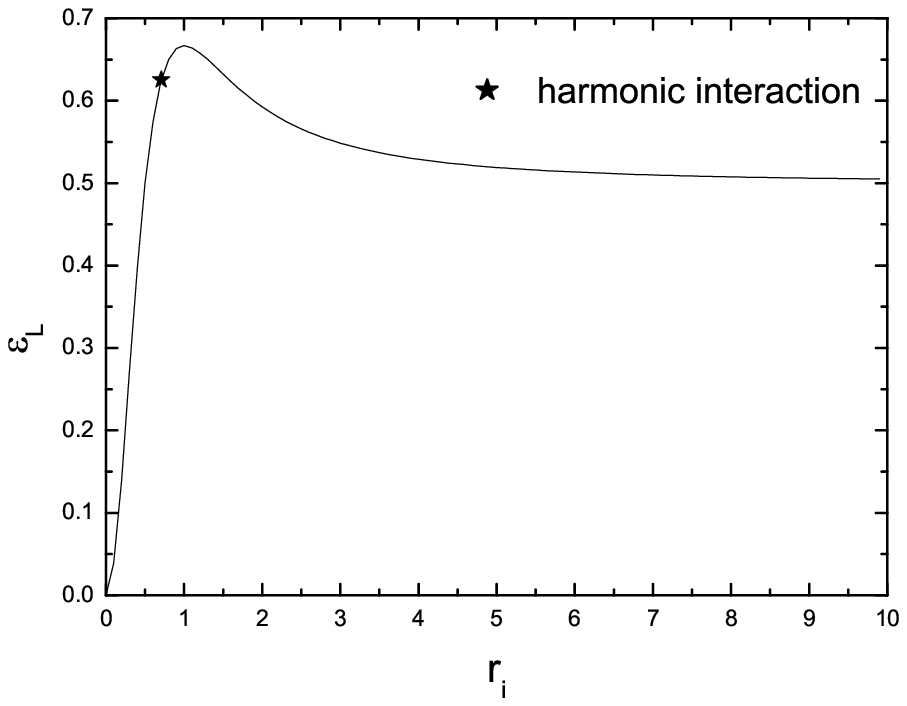}
\vspace{0.5cm}
\includegraphics[scale=0.75,angle=0]{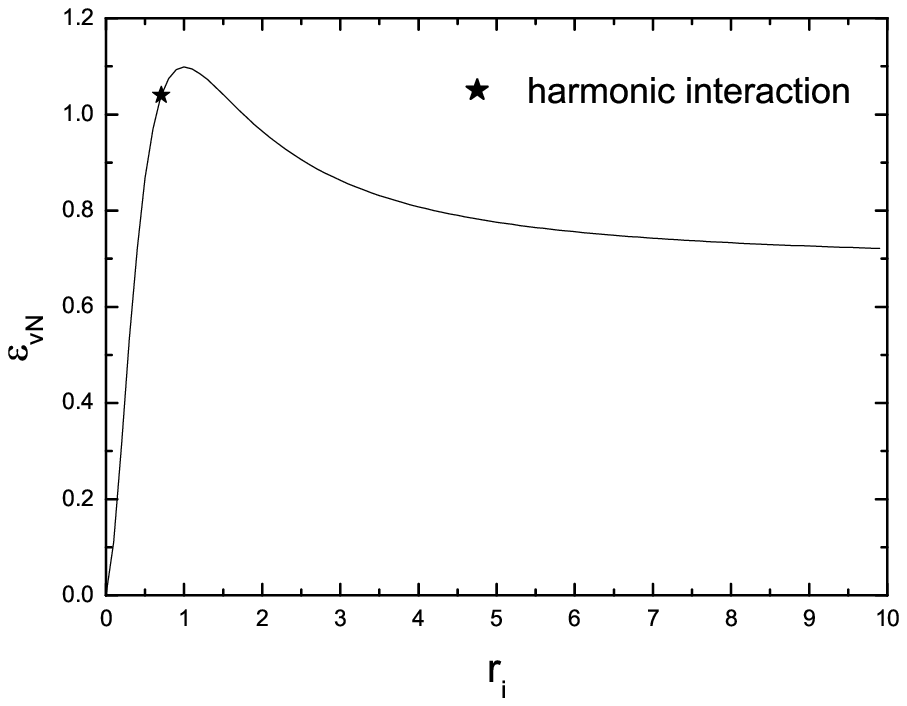}
\caption{$\varepsilon_L$ (left panel) and $\varepsilon_{vN}$
(right panel) as a function of $r_i$ for the case $n_1+n_2=2$ and
an arbitrary interaction $V(x_1-x_2)$. The stars in both plots correspond
to the entanglement amount for the Moshinsky atom (harmonic
interaction). All depicted quantities are
dimensionless.\label{figu_2}}
\end{center}
\end{figure}

The corresponding normalized eingevectors can be expressed as
follows
\begin{eqnarray}
 |\psi_1'\rangle&=&|\psi_1\rangle\nonumber\\
 |\psi_2'\rangle&=&|\psi_4\rangle\nonumber\\
 |\psi_3'\rangle&=&\frac{1}{\sqrt{2}}(|\psi_2\rangle+|\psi_3\rangle)\nonumber\\
 |\psi_4'\rangle&=&\frac{1}{\sqrt{2r_1^2+1}}(-r_1|\psi_2\rangle+r_1|\psi_3\rangle+|\psi_5\rangle)\nonumber\\
 |\psi_5'\rangle&=&\frac{1}{\sqrt{2r_2^2+1}}(-r_2|\psi_2\rangle+r_2|\psi_3\rangle+|\psi_5\rangle),
\end{eqnarray}
where $r_1$ and $r_2$ are functions of the matrix 
elements $b,c,d,e$ given by the eexpressions,
\begin{equation}
 r_1 = \frac{C_1+cR}{C_2+dR}, \,\,\,
 r_2 = \frac{C_1-cR}{C_2-dR},
\end{equation}
with
\begin{equation}
 C_1 = -bc+c^2+2d^2+ce, \,\,\,
 C_2 = d(b+3c-e),
\end{equation}
and
\begin{equation}
 R=\sqrt{b^2+c^2+8d^2+2ce+e^2-2b(c+e)}
\end{equation}
Now we calculate the amounts of entanglement of the states $|\psi_4'\rangle$
and $|\psi_5'\rangle$ that depend (in the same way) on $r_1$ and $r_2$
respectively. The values adopted by these two constants depend
on the form of the interaction $V(x_1-x_2)$. The general expression
for the amount of entanglement of the states $|\psi_4'\rangle $
and $|\psi_5'\rangle $ are,
\begin{equation}
 \varepsilon_L(r_i)=1-\frac{2r_i^4+1}{(2r_i^2+1)^2} \,\,\,\,\,\,\,\,\,\,i=1,2
\end{equation}
\begin{equation}
 \varepsilon_{vN}(r_i)=\ln(2r_i^2+1)-\frac{4r_i^2}{2r_i^2+1}\ln r_i \,\,\,\,\,\,\,\,\,\,i=1,2
\end{equation}
We plot these expressions in Fig. \ref{figu_2}. Note that the
particular value $r_i=\frac{1}{\sqrt{2}}$ corresponds to
the harmonic interaction in the Moshinky model.

\section{Conclusions}

 By recourse to a perturbative approach we studied the 
 entanglement-related properties of a system consisting of two 
 interacting spin-$\frac{1}{2}$ fermions (``electrons") confined 
 by an external potential. Our present results clarify some aspects 
 of the basic entanglement features exhibited by particular two-electrons
 models studied previously, and shed some light upon the fact that these 
 systems share important qualitative entanglement properties that 
 are also observed in more general cases. Our analysis highlights 
 the important role played 
 by the degeneracy of the energy levels of the ``unperturbed"
 (interaction-free) Hamiltonian $H_0$. The non-vanishing
 entanglement exhibited by the interacting particles in the limit
 of vanishing interaction is due to the particular eigenbasis of $H_0$
 ``chosen" by the interaction. This amount of entanglement tends
 to increase with the alluded degeneracy which, in turn, tends to
 increase with energy. {\it This sheds light on the physical reasons behind
 the fact (observed in all cases studied so far) that the amount of
 entanglement exhibited by the eigenstates of two-electrons systems tends
 to increase with energy}. These basic trends do not depend on the
 particular entanglement measure employed, as has been shown in the 
present work, where entanglement measures based upon the linear and
the von Neumann entropies were considered. In connection with this
 point it is worth to mention a relevant question raised in a recent 
 review article on entanglement in atoms and molecules \cite{TMB10}: 
 does the existence, for some excited states, of a finite amount of 
 entanglement in the limit of vanishing interaction depend upon the
 particular entanglement measure employed? As already mentioned, the
 answer to this question is that the alluded feature constitutes 
 an intrinsic property of the systems under consideration, which does 
 not depend on the entanglement measure used. \\

 As particular illustrations of the above considerations we have

 \begin{itemize}
 \item{Computed for general confining and interaction potentials, 
 in the limit $\lambda \rightarrow 0$, the entanglement measures based 
 upon the linear entropy and the von Numann entropy of the excited states 
 associated with the four-fold degenerate unperturbed first excited state.}
\item{Obtained for a harmonic confining potential and a generic interaction
potential $V(x_1-x_2)$, the entanglement of the eigenstates corresponding 
in the above limit to the second excited unperturbed energy level (given by 
$n_1+n_2=2$).}
\item{Determined, for a harmonic confining potential and an arbitrary 
interaction, upper bounds on the amounts of entanglement exhibited 
by the system's eigenstates in the limit of vanishing interaction. 
These upper bounds are expressed in terms of the quantum numbers 
$n_1$ and $n_2$.}
\end{itemize}

 The results advanced here corresponding to the entanglement in the
 limit of vanishing interaction are {\it exact}, and our procedure 
 can in principle be applied to any excited state of systems of the 
 kind considered in this work. In this sense, the perturbative method 
 used here is not (in intself) the fundamental protagonist of our 
 present considerations. The perturbative approach was used only as 
 a tool to determine (exactly) the entanglement features of
 the system's eigenstates in the limit of vanishing interaction,
 in order to get some insight on the qualitative entanglement
 features of two-elctrons models. It is worth stressing that the 
 entanglement exhibited by these systems 
 in the limit $\lambda \rightarrow 0$ constitutes a basic, 
 dominant aspect of their entanglement-related characteristics.
 It is to be expected, on  general physical grounds, that the 
 entanglement-degeneracy relationship uncovered here constitutes a 
 typical feature of atomic models. This provides a first step towards 
 explaining the fact that the general entanglement features exhibited 
 by soluble models such as the Moshinsky one are also observed in other
 two-electrons systems. 

 The ideas discussed here may also be useful for the analysis of 
 entanglement-related aspects of other scenarios involving interacting 
 fermions, such as those appearing in molecular or solid state physics.
 Just to mention one example, a situation similar to the one observed in 
 the atomic models considered here also occurs when studying the behaviour 
 of electronic entanglement in the dissociation process of diatomic 
 molecules \cite{Esqui11}. One observes that, for instance, in the limit of 
 large values of the reaction coordinate (corresponding to vanishing interaction 
 between the electrons in the system) describing the dissociation of $H_2$, the 
 electronic entanglement does not tend to zero  \cite{Esqui11}.

\acknowledgments This work was partially supported by the Projects
FQM-2445 and FQM-207 of the Junta de Andalucia and the grant
FIS2011-24540 of the Ministerio de Innovaci\'on y Ciencia (Spain).
A.P.M. acknowledges support by GENIL trhough YTR-GENIL Program.

\end{document}